\documentstyle[11pt,fleqn,epsf]{article}
\def\ref{par\noindent\hangindent=6mm\hangafter=1}
\begin{document}
\vbox{
\rightline{Nuovo Cimento B 109, 423-430 (April 1994)}
\rightline{physics/9711015}
}
\baselineskip 8mm
\begin{center}
{\bf  On the ``circular vacuum noise" in electron storage rings}

\bigskip

Haret C. Rosu\footnote{e-mail: rosu@ifug3.ugto.mx}\\ 
Instituto de F\'{\i}sica de la Universidad de Guanajuato,
Apartado Postal E-143, Le\'on, Guanajuato, M\'exico

\end{center}

\bigskip
\bigskip

\begin{abstract}

We discuss the proposal of Bell and Leinaas (BL)
to measure the circular Unruh effect in storage rings.
The ideal concept `circular Unruh effect' has a more realistic
correspondent such as `circular vacuum noise' used by Shin Takagi
[Prog. Theor. Phys. Suppl. 88, 1 (1986)]
The BL resonance
behavior does not fit to the SPEAR first order betatron resonance at
3.605 GeV, but of course, the real experimental situation is much more
complicated, corresponding, as a matter of fact, to an even more
general term that one may call `synchrotron noise'.
In the final section we focus on radiometric aspects of
storage ring physics, such as the problem of establishing better
quantum field radiometric standards at high energies. The Unruh-like
effect could be a useful guide for that purpose.

\end{abstract}


\vskip 2cm

\section{ Introduction}  

About thirty years ago Sokolov and Ternov showed that electrons
circulating in a vertical and uniform magnetic field get polarized
 \cite{st}.
This occurs because in the quantum synchrotron emissions there is a
very small spin-flip power( only $10^{-11}$ of the classical continuous
power), which is accumulating over a timescale of tens of minutes to a
few hours to
 give a final asymptotic polarization
$P_{lim}=8/(5\sqrt{3})= 0.924$. This number found three decades ago as the
result of a
simple theoretical QED exercise is today a famous figure of accelerator
physics. In the seventies, Derbenev and Kondratenko \cite{dk}
obtained a
more realistic formula for the limiting beam polarization containing
the  spin-orbit coupling function. This vector parameter
 expresses the
correlations among the positions of the synchrotron emission events on
the orbit and the spin motion (fluctuations of the precession axes).
It is completely determined by the
magnetic lattice of the accelerator. In a certain sense we could say
that the synchrotron vacuum fluctuations are angularly constrained
 by the
magnetic boundaries of the guiding structure via the spin axis of the
electrons (a kind of magnetic orientational Casimir effect).

The Sokolov-Ternov effect has been confirmed at the major
synchrotrons and storage rings of the world. Even at LEP,
the largest storage ring at the moment, a Compton polarimeter
 detected
some polarization \cite{k}. Also transverse beam
polarizations have been measured at HERA by means of a laser polarimeter
\cite{Barp}.
However, the whole spin game becomes technically interesting because of
 the depolarizing resonances (DRs) that one encounters easily
 at any high energy circular accelerator. In a way, accelerator DRs
  are similar
 to the classical resonances of the Solar System, asteroid belt and
 planetary rings, of course at much different scales of time and
 space.
The accelerator DR's are a consequence of the anomalous (presumably
 irrational)
  magnetic moment of the electron and the proton ($a=\frac
 {g-2}{2}$; $a_{e}= 0.00116$, $a_{p}=1.793$). DRs have been classified
 according to their cause and their parameters have been determined
 with great accuracy. The point is that at high energies the density
 of DRs becomes embarassingly large, a fact of great concern for the
 accelerator physicists.

\section { Can one see the BL effect ?}   

   The DK polarization formula of 1973 is overwhelmingly used
 by accelerator people. It includes in a well-established way the
  effects of DRs on the equilibrium polarization.
  This formula reads
  $$P_{eq}^{DK}=\frac{8}{5\sqrt{3}}
\frac{<\mid {\rho} \mid ^{-3}\hat {b} (\hat {n}- F_{DK})>}
{<\mid {\rho} \mid ^{-3}(1-2/9(\hat {n}\cdot \hat {v}) ^{2}+11/18
\mid F_{DK} \mid ^{2})>}~,      \eqno(1)$$
where $F_{DK}=\gamma\frac{\partial \hat {n}}{\partial \gamma}$
is the spin-orbit coupling function, which takes into account the
depolarizing effects of jumps between various trajectories differing
from the reference closed orbit, $\rho$ is the bending radius, $\hat {b}$
a unit vector along the transverse magnetic field component. The
brackets indicate an average over the ring circumference and over
the ensemble of particles in the beam. The unit vector $\hat {n}$ is the
time-independent spin solution of the BMT equation, attached to each
particle trajectory.

 On the other hand, motivated by Unruh
 thermal-like effect, which can show up in circular motion too
 (see section 3),
 Bell and Leinaas \cite{bl} developed a formalism which is
  closer to quantum
 field theory and therefore more acceptable by theoretical physicists.
 Aiming at very peculiar effects, first of all, the
 correct way of taking into account vertical electron recoils,
  Bell and Leinaas have been forced to consider the case when the
spin-orbit function
  is zero. This is valid only for perfectly aligned weak focusing
  storage rings for which the accelerating fields are independent of
  arc length and the magnetic field is vertical on the perfectly planar
   closed orbit with a small magnetic gradient
   $n=-(B/R)^{-1}(\partial B/\partial r)$. The meaning of $r$ is the
   radial displacement from the closed orbit R. Usually the vertical
   betatron fluctuations as determined by the horizontal synchrotron
   emissions are a negligible effect as compared to the more common
   accelerator stochastic excitations. However, for weak focusing
    machines such effects expressed by Bell and Leinaas in terms of a
    parameter $f$ have also a linear contribution to the limiting
    polarization and become predominant over the main part of the
    stochastic excitations included in the spin-orbit function
     of DK.
    Moreover, in an ideal perfectly aligned machine the DK spin-orbit
    function is actually zero.
The corresponding equilibrium betatron emittance of the
    vertical oscillations is calculated by BL explicitly by means
    of the standard Lorentz-Dirac equation in order to take into
    account the radiation damping (on the Langevin character of the
 Lorentz-Dirac equation we shall comment in a future work, \cite{r}). 
The final BL polarization formula is 
$$ P_{BL}=(8/5\sqrt{3})\times\frac{1-f/6}{1-f/18+ 13f^{2}/360}~,
 \eqno(2)$$
where the BL parameter f is given by
$$f=\frac{2}{\gamma}\times \frac{\nu_{s} Q_{\beta}^{2}}
    {Q^{2}_{\beta}-\nu^{2}_{s}}~.  \eqno(3)       $$
In Eq.(3) $\gamma $ is the relativistic kinematical factor,
$\nu_{s}$ is the spin tune, ($\nu_{s}=a_{e}\gamma = E/0.441 (GeV)$),
and $Q^{2}_{\beta}$ is the betatron tune.

The BL formula has the usual intrinsic betatron resonance condition
$\nu_{s}=Q_{\beta}= $ \( \sqrt{n} \). However the $f$ parameter implies a
more intricate behavior of the polarization in passing through the
resonance, which should be isolated and of first order, namely
$P_{eq}$ drops from 0.924 to -0.169, then increases to 0.992 to fall
eventually to 0.924.
A thorough comparison of the DK and BL formalisms have been provided
by Barber and Mane \cite{bm}. At the same time Mane has generalized
the BL result to the case of strong-focusing storage rings (see
formula (41) in ref. \cite{bm}). The $f$
parameter turns now to be a vector quantity defined as
${\bf f}=-(2/\gamma)\partial{\bf n}/\partial\beta_{b}$ where again the
subscript $b$ denotes
the direction of the magnetic field, not necessarily vertical.
This allows a coupling of the horizontal betatron and
synchrotron oscillations to the vertical fluctuations;
{\bf n} is the chosen spin quantization axis.
The only first order resonance at which one might think of a BL effect
on the SPEAR polarization data \cite{j}, is the resonance
 $\nu_{s}=3+\nu_{y}$ at
E= 3.605 GeV, but we have to emphasize that the experimental conditions
are very different from those required by the ideal BL case.
The SPEAR E= 3.605 GeV resonance would correspond to n=67. This is
 already
a rather high n to make the BL effect unobservable. At SPEAR energies we
encounter a well developed forest of resonances not the isolated BL
situation.Also SPEAR
has a superperiodicity of two for which all the odd resonances are
 forbidden. Their presence is due only to higher order effects, making
 them very narrow, (for an interpretation in terms of nonlinear
  tunespreads see \cite{m1}, \cite{kw}).

\section{ More about circular vacuum noise}             

 We now come to the problem of interpretation. Bell and
Leinaas were motivated in their particular treatment of fluctuations
in electron storage rings by the chance of revealing Unruh effect of
vacuum fluctuations. As it is well known there is a close parallel
between Hawking effect and Unruh effect \cite{r1}.
Their proposal was chronologically the second one to detect such
fundamental effects within terrestrial laboratories after that of
Unruh \cite{u}, who developed a hydrodynamical analogy for
Hawking effect.
It has been considered as the most feasible one for a long time.
However,
in their 1987 paper BL are aware of the difficulty of introducing a
temperature parameter for synchrotron fluctuations. We recall that
for one parameter problems, like Schwarzschild black holes and Rindler
linear accelerated motion, the quantum field vacuum turns formally into
an equilibrium thermal state. One may introduce a thermodynamic
temperature directly related to that one parameter of the problem,
the Schwarzschild mass in the first case, the constant proper
acceleration in the latter. There are papers connecting the usual QED
bremsstrahlung and zero-energy Rindler photons \cite{HMS}.

However, the situation is much different in the uniform circular motion,
and not only because there is no horizon, but more important because the
acceleration, despite being a constant, has the rather peculiar form
given by
$$\alpha_{c}=\frac{\rho\omega_{0}^{2}}{1-\rho^{2}\omega_{0}^{2}}~.
 \eqno(4)$$
Here $\rho$ is the radius of the orbit and $\omega_{0}$ is the
cyclotron frequency. Units such that $\hbar =c=1$ are used.

The point is that in the circular motion the vacuum energy density
cannot be written in the canonical Planck spectrum form \cite{ksy}.
Since the Fourier transforms of the Wightman functions have
 singularities of the branch cut type (implying $\Theta$ functions)
 one will find out an energy density of the vacuum fluctuations
 as a sum over cyclotron harmonics. The quantum zero-point noise is
 in this case of an intermediate nature between an additive and
 a multiplicative noise. We shall
 address this very interesting problem in another work \cite{r}.

 The vacuum energy density of a massless scalar field
 in the circular case can be written as follows \cite{ksy}
 $$\frac{d e}{d \omega}=\frac{\omega^{3}}{\pi^{2}} \{ 1/2 +
 \frac{\omega_{0}}{\gamma \omega} \sum_{n=0}^{n=\infty}
  \frac{\beta^{2n}}{2n+1}\sum_{k=0}^{n} (-1)^{k}\frac{n-k-
  \frac{\omega}{\gamma \omega_{0}}}{k!(2n-k)!}\Theta(n-k-
  \frac{\omega}{\gamma \omega_{0}})\}~.    \eqno(5)   $$
From Eq.(5) one can see that to a certain power of the velocity
many vacuum cyclotron harmonics could contribute making
the energy density spectrum a quasi-continuous one
 but with
different shape and scale as compared with a pure Planckian spectrum.

The na\"{\i}ve introduction of the circular-motion acceleration
$\alpha_{c}=\gamma^{2} \omega_{0} v$ into the Planckian function
$\{ \exp{[2\pi \frac{\omega v}{\gamma^{2} \omega_{0}}]-1}\}^{-1} $
shows an essential singularity at v=0 which does not allow an
expansion in velocity powers.

Hacyan and Sarmiento \cite{hs} developed a formalism, very close
to the scalar case, for calculating the vacuum stress-energy tensor
of the electromagnetic field in an arbitrarily moving frame and applied
it to a system in uniform rotation. They provided formulas for energy density, 
Poynting flux, and stress of the zero-point field in such a frame. Moreover, 
Mane \cite{m2} has suggested the Poynting flux of Hacyan and Sarmiento to be
in fact synchrotron radiation when coupled to an electron.

The relationship of circular noise and Rindler noise has been
under focus in the works of Letaw \cite{l}, Gerlach \cite{g},
and Takagi \cite{t}. These authors managed to determine an exact
connection for the massless scalar field between the circular noise
and so-called ``drifted Rindler noise" seen by a quantum detector
which is uniformly accelerated but has also a constant speed in the
 direction perpendicular to the acceleration. It is just the limiting
 case of infinite radius for circular uniform motion. In the three-
 dimensional Minkowski space the circular motion is a helix winding
  around the time axis. The Rindler trajectory is also a helix
  this time winding with an imaginary pitch around a space axis.
  They are examples of the six classes of stationary motions
  obtained by Letaw \cite{l} and possessing stationary noises.
  Nevertheless these stationary noises do not satisfy the KMS
  condition, i.e., the principle of detailed balance in quantum field
  theory \cite{r}. To quote
Takagi \cite{t}, ``the effective temperature depends on the energy
  and does not provide a very useful concept."

\section{ On radiometry at storage rings}          

The considerations in the previous section make us expound a
little on the (primary) radiometric standards at high energies.
As it is well known the common blackbody radiators are limited to
$3\times 10^{3}$ K for technical reasons. The idea to use electron
storage rings with their magnetobrehmsstrahlung/synchrotron spectrum as
 primary
radiometric standards at high energies is not new, and important
contributions have been made in the past \cite{ku}. Already 4
 electron
storage rings working at 800 MeV, approximately,
are used as radiometric standard sources: SURF II, VEPP 2M,
TERAS, and BESSY. The
 projected BESSY II will also be designed with regard to radiometric
 applications \cite{te}. The primary radiometric character of the
 synchrotron spectrum is based on the famous spectral photon flux
  formula of Schwinger \cite{sch}.

If the comparison between the (non-thermal) magnetobrehmsstrahlung
 spectral
distribution and the (thermal) black body one is made by identifying
their two maxima \cite{tm}, one concludes that for a
beam of 1 GeV the ``temperature" of the maximum of the brehmsstrahlung
(synchrotron) radiation is about $10^{7}$ K. It is a gain  of 4 orders
of magnitude for the important field of radiometry. Nevertheless,
as has been argued in section 3 above, we are not
in completely equivalent situations.The shapes of the two spectral
densities are different, i.e.,
$$ f(y)= (9\sqrt{3} /8\pi) y \int_{y}^{\infty} K_{5/3}(x)dx  \eqno(6) $$
for the synchrotron radiation, as compared to 
$$ \phi(y)= (15/\pi^{4}) \frac{y^{3}}{e^{-y}-1}   \eqno(7)  $$
for the blackbody radiation.
In Eqs.(6) and (7) the $y=\frac{\omega}{\omega_{max}}$ variable is a scaled
frequency,
where $\omega_{max}$ is corresponding to the maximum
of the synchrotron radiation $\omega_{max}= \omega_{c} \gamma^{3} $,
$\omega_{c}$ being the cyclotron angular frequency.

One may hope to obtain non-thermal distributions from a thermal one
by means of q-deformations \cite{md}. Indeed, the features of the
emitted spectrum are strongly dependent on the electron-photon
interactions along the beam trajectory and a q-boson interpretation
seems natural.

In the search for a better primary radiometric standard we recall that
a formal truly thermal ambience such as the Unruh one is not a
local property of the trajectory, but a global one \cite{phy}.
We need a constant power spectrum uniformly distributed over
the radiation spectrum of the electron. To get this
an undulator of a special type made of a
collection of so-called ``short magnets" \cite{ba} is needed. For
a single ``short magnet", it has been shown \cite{ba} that the electron will
radiate a white noise with the angular frequencies
 distributed from zero up to the frequency
 $ \omega_{short}=( c/l) \gamma^{2}$.
 Here $l = 2\alpha R$ is the arclength of the electron trajectory in the
 ``short magnet". The criterion for a magnet to be a ``short magnet" is
 $\alpha \ll \frac{mc^{2}}{E}$.
They shift the maximum of the synchrotron
radiation to longer wavelengths , and depending on their number we
could move the maximum of the synchrotron spectrum to whatever
wavelength we like in the spectral width of the quasi-white noise.

 It seems therefore that various types of insertion devices will be of
  great help in the field of radiometry \cite{mu}. Especially for storage
  rings working at higher energies (ESRF, Spring 8, APS) detailed models for
  the radiation at the insertion devices are required for its use in
  radiometry.

A highly interesting field regarding the connections between
the coherence and polarization features of the insertion devices
 and radiometry is foreseeable (see \cite{mu}
 for a first step in this direction).
 It is known that even such a good simulator
of blackbody radiation as cavity radiation has a spatial frequency
spectrum, which is not white over the real plane waves \cite{ma}.

It is worth noting also that high-energy radiometry could be done in
 the future for other types
of brehmsstrahlungs, e.g., the beamstrahlung at linear colliders.

Radiation noises and radiometry at colliders are obviously connected
with chaos problems. Because in the phase space of a system possesing
N dof's the uncertainty principle replaces every portion
of the continuum of classical trajectories of states in each
$\hbar ^{N}$ volume cell by only one, so-called quantum state, one
immediately concludes that the physical quantum chaos is more ordered
than the classical (geometric) one. The latter is only a limiting
pure mathematical chaos from the point of view of real, physical
measurement. But if one will insist to go beyond quantum theory of
physical measurements a knowledge of the topology and geometry of
the proximity of points is unavoidable \cite{l}.

\section{Concluding Remarks}                    

We presented a discussion of the `circular Unruh' effect put forth by
Bell and Leinass \cite{bl}. Recently, Cai, Lloyd, and Papini \cite{CLP}
claimed that the spin-rotation-gravity coupling, also known as
Mashhoon effect in gravitation theory \cite{M}, is practically stronger
than the acceleration effect at all available energies, but the comparison
is rather difficult and not so direct.
Another interesting effect is mentioned by Fr\"{o}hlich and Studer \cite{FS}
for the case of non-electronic storage rings. A beam of non-relativistic
particles with spin  display a variant of the Einstein-de Haas effect which
could show up as a tidal Zeeman energy affecting, after relaxing to a
steady state, the ratio of spin-up to spin-down ions in the beam.

Finally, we sketched above some radiometric aspects of storage ring physics
that may be useful for future detailed analyses.

\section*{ Acknowledgments}
This work was partially supported by CONACyT Grant No. F246-E9207.

The author would like to thank Professor Abdus Salam, the International
Atomic Energy Agency and UNESCO for hospitality at the International
Centre for Theoretical Physics, Trieste, where this work has
been started.

The author is grateful to Drs. S.R. Mane and D.P. Barber
for helpful comments.


\end{document}